\documentclass[preprint]{aastex62}
\usepackage{gensymb}
\usepackage{soul}
\usepackage{xcolor}
\usepackage{ulem}

\begin{document}


\title{29P/Schwassmann--Wachmann 1, A Centaur in the Gateway to the Jupiter-Family Comets}

\correspondingauthor{G. Sarid}
\email{galahead@gmail.com}
\author[0000-0001-5678-5044]{G. Sarid}
\affil{University of Central Florida, Florida Space Institute, USA}

\author[0000-0001-8736-236X]{K. Volk}
\affil{University of Arizona, Lunar and Planetary Laboratory, USA}

\author[0000-0002-1717-2226]{J.K. Steckloff}
\affil{Planetary Science Institute, USA}
\affil{University of Texas at Austin, Department of Aerospace Engineering and Engineering Mechanics}

\author[0000-0002-8378-4503]{W. Harris}
\affil{University of Arizona, Lunar and Planetary Laboratory, USA}

\author[0000-0003-4659-8653]{M. Womack}
\affil{University of Central Florida, Florida Space Institute, USA}

\author[0000-0002-0004-7381]{L.M. Woodney}
\affil{California State University San Bernardino, Department of Physics, USA}


\begin{abstract}

Jupiter-family comets (JFCs) are the evolutionary products of trans-Neptunian objects (TNOs) that evolve through the giant planet region as Centaurs and into the inner solar system. 
Through numerical orbital evolution calculations following a large number of TNO test particles that enter the Centaur population, we have identified a short-lived dynamical Gateway, a temporary low-eccentricity region exterior to Jupiter through which the majority of JFCs pass. We apply an observationally based size distribution function to the known Centaur population and obtain an estimated Gateway region population. We then apply an empirical fading law to the rate of incoming JFCs implied by the the Gateway region residence times. Our derived estimates are consistent with observed population numbers for the JFC and Gateway populations. Currently, the most notable occupant of the Gateway region is 29P/Schwassmann--Wachmann 1 (SW1), a highly active, regularly outbursting Centaur.
SW1's present-day, very-low-eccentricity orbit was established after a 1975 Jupiter conjunction
and will persist until a 2038 Jupiter conjunction doubles its eccentricity and pushes its semi-major axis out to its current aphelion. Subsequent evolution will likely drive SW1's orbit out of the Gateway region, perhaps becoming one of the largest JFCs in recorded history. 
The JFC Gateway region coincides with a heliocentric distance range where the activity of observed cometary bodies increases significantly. SW1's activity may be typical of the early evolutionary processing experienced by most JFCs. Thus, the Gateway region, and its most notable occupant SW1, are critical to both the dynamical and physical transition between Centaurs and JFCs.

\end{abstract}

\keywords{Centaurs; Short period comets; Trans-Neptunian objects; Kuiper belt; Solar system; Orbits; }


\section{Introduction} \label{sec:intro}

Centaurs are a transient population of icy bodies that dynamically
link the outer solar system's trans-Neptunian object (TNO) population with the Jupiter-family comets (JFCs).  
Centaurs have perihelia and semi-major axes between the orbits of Jupiter and Neptune \citep{Jewitt2009}, with frequent gravitational perturbations from the giant planets driving their dynamical evolution. JFCs are comets whose dynamics are controlled by Jupiter. They are often defined using the Tisserand parameter with respect to Jupiter ($T_J=a_J/a+2\sqrt{a/a_j(1-e^2)}\cos~i$) to have $2<T_J<3$. The Tisserand parameter neglects the presence of other planets, so we apply another definition for JFCs: objects with perihelia interior to Jupiter ($q<5.2$~au) and aphelia well-separated from Saturn ($Q<7$~au). We refer to this as our JFC [q-Q] definition. 

\begin{figure}
\centering
\includegraphics[width=3.5in]{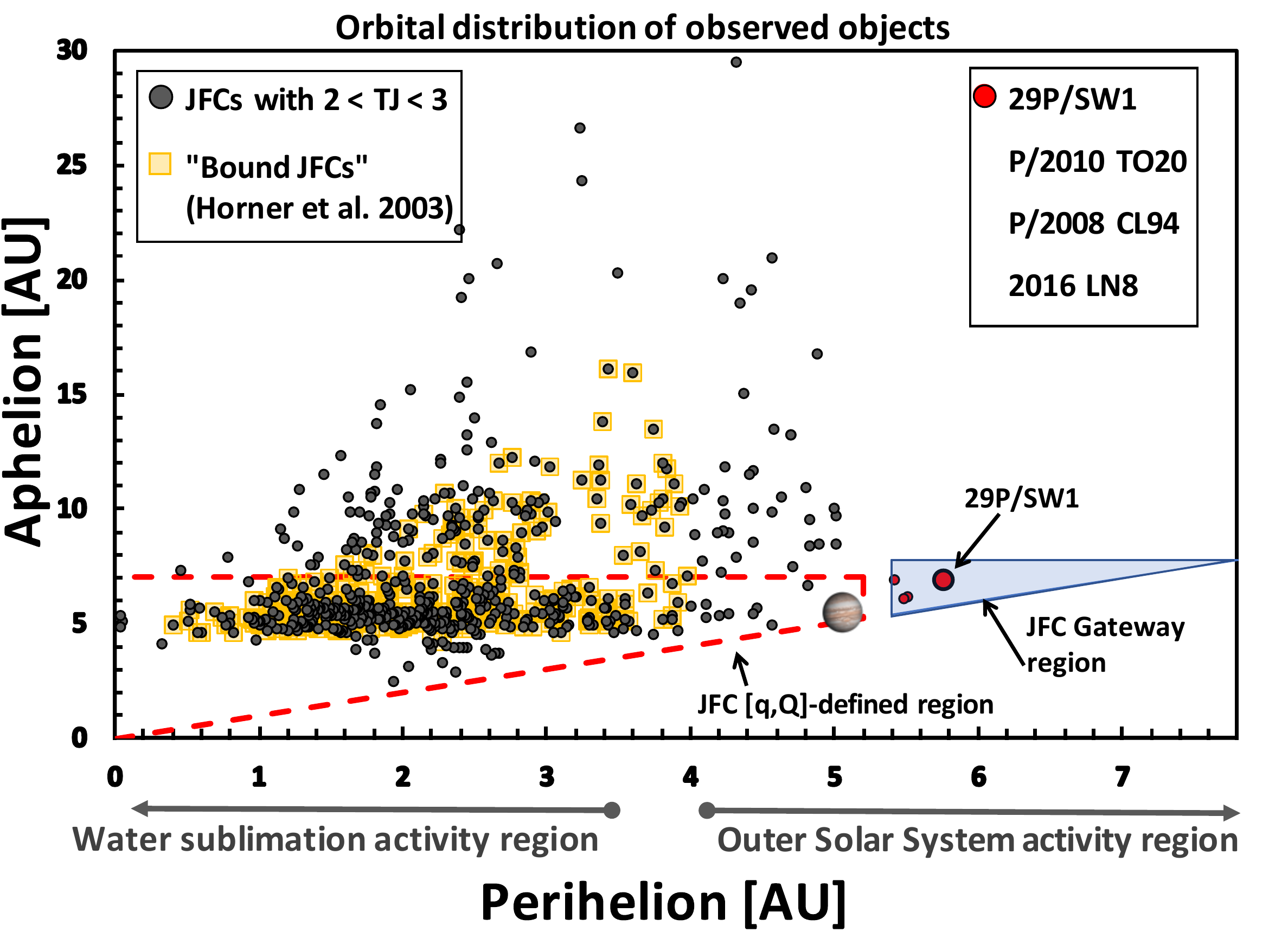} 
\caption{Perihelion-aphelion distribution of JFCs (gray circles; $2<T_J<3$) and several objects of interest, including SW1 (red circles). Our [q,Q]-defined JFC population is enclosed by the dashed red line. The ``JFC Gateway region'' is shown in blue (see Section~\ref{sec:methods}). A Tisserand parameter-based classification of ``bound JFCs''  \citep{Horner2003} is shown for comparison (yellow squares). We note a distinction in the main mechanisms driving activity using gray arrows along the x-axis: the water--ice sublimation controlled region extends inward from $\sim3.5$~au and mechanisms other than water sublimation are prominent from $\sim4$~au outward.  
}
\label{f:definitions}
\end{figure}

Fig.~\ref{f:definitions} illustrates the JFC population defined using $T_J$ and [q-Q]. For reference, we include another dynamical mapping of JFCs that divides objects with $T_J<3$ into several dynamical groups (e.g., `loosely bound' $2.5<T_J<2.8$ and `tightly bound' $2.8<T_J<3$), all with $q<4$~au \citep{Horner2003}. Our [q-Q]-defined JFCs include $\sim75\%$ of those `bound' objects while also including some objects with $q>4$ that may be transitioning to JFC status.

Centaur evolution is dominated by chaotic gravitational perturbations, so an individual Centaur's orbital pathway through the giant planet region is highly sensitive to its initial conditions. Nevertheless, trends emerge in the evolution of Centaurs as a population. Passage through the Centaur region typically takes $\sim$1--10 Myr \citep{Tiscareno2003,DuncanEtAl2004,Disisto2007}, with multiple possible outcomes. 
The giant planets can eject Centaurs back into the TNO's scattering population, into the Oort Cloud, or out of the solar system entirely \citep[see, e.g.,][]{DonesEtAl2015}. Centaurs that avoid such ejection evolve inward into the JFC population, passing through many gravitational interactions, and conceivably a transition region to enter (or exit) the inner solar system. The JFC population is observed to be prevalent with small nuclei, which is consistent with a total Centaur population of order $10^7$ objects larger than 2~km in diameter \citep{Sheppard2000}.

29P/Schwassmann--Wachmann 1 (hereafter SW1), discovered while outbursting in 1927, was the first observed small body with an orbit entirely beyond Jupiter. While its activity and outburst characteristics led to its classification as a comet, the subsequent discovery of other bodies between Jupiter and Neptune, followed by dynamical models linking TNOs and JFCs, place SW1 in better context as a member of the Centaur population. SW1's low-inclination ($i=9.37\degree$) and nearly circular orbit (heliocentric distance range 5.77-6.28~au; $e=0.043$) put it at the inner edge of the Centaur region; its Tisserand parameter ($T_{J}=2.985$) also places it at the cusp between the two populations \citep[e.g.][]{Fernandez2018}.  

While initially SW1 appears to be an outlier, there are now three other observed objects in its orbital region: P/2010~TO20 LINEAR-Grauer, P/2008~CL94 Lemmon and 2016~LN8. 
Fig.~\ref{f:definitions} shows these objects, highlighting their position relative to the JFCs. Their diameters are estimated to be between 4 and 12~km, from coma-corrected nucleus photometric observations \citep[][for P/2010~TO20 and P/2008~CL94, respectively]{Lacerda:2013, Kulyk:2016} or {\textit H}-magnitudes (MPC database for 2016~LN8, assuming $10\%$ albedo). For centaurs, an ensemble analysis gives a mean albedo of $8\% \pm 4\%$ \citep{Stansberry2008, Bauer2013}, which is consistent with diameters between $\sim$10 and 14~km. NEOWISE provided an independent diameter estimate for P/2010~TO20 of $10.6$~km \citep[][see online data]{Bauer2017}.

These objects meet the orbital definition of a Centaur, but they also have $T_J$ near the upper boundary of the JFC definition (with 2016~LN8's $T_J$ in the middle of the JFC range due to its 
high inclination of $43^\circ$). Interestingly, in 2011, P/2010~TO20 had an orbit confined between 5.1 and 6.1~au, compared with its current orbit spanning 5.5--6.1~au;
this short-timescale variation underscores the transient nature of this Gateway region between Centaurs and JFCs (see the definition of ``JFC Gateway Region'' in Sec.~\ref{sec:methods}).

SW1 is the only continuously active, large Centaur in the Gateway region. Additionally, it is the only icy object with a predictable pattern of major outbursts \citep{Trigo2008,Miles2016}, despite its orbit being entirely beyond the region where surface water--ice sublimates efficiently (distances $\lesssim$3--4~au). However, the Gateway region where SW1 resides (heliocentric distances of $\sim$5--7~au) is where distantly active comets tend to show strong increases in outgassing and outbursts. The rapid crystallization of amorphous water ice could drive this activity \citep[e.g.][]{SaridPrialnik2009, WomackDistantReview2017}. Alternatively, newly exposed volatile ice patches (e.g., CO, CO$_2$) or release of sub-surface volatile gas pockets could also drive such distant activity \citep[see, e.g.][]{Prialnik:2004CometsII,Guilbert-Lepoutre2015}. 

The first direct millimeter-wavelength detection of CO emission in any cometary object was made in SW1 by \citet{senjewitt94IAU, SenayJewitt94Nature}. Similarly, over a span of 18 months, SW1's CO production rate was noted by the same observers to range between 1.1 and 1.4$\times10^{28}$~molecules/sec, with a double-peaked velocity profile (0.5 to -0.1km) that was interpreted as a jet \citep{SenayJewitt95DPS}. Interestingly, this supports a more current model for outbursts, based on a combination of observation analyses \citep{Schambeau2018}. 
CO gas has been measured in SW1's coma during all activity phases \citep{crovisier95,fes01,gun08,paganini2013}, even sometimes exceeding CO production rates of comet C/1995 O1 (Hale--Bopp) at comparable heliocentric distances \citep[see][]{WomackDistantReview2017}. Simultaneous observations of gas and dust during times of high activity could provide further observational constraints on these outburst mechanisms (K. Wierzchos \& M. Womack 2019, in preparation).

It may be the case that the Gateway region, with its potentially overlapping activity mechanisms, is where inbound JFCs experience the onset of bona fide cometary activity. SW1's orbital configuration and unique activity characteristics motivate us to examine its orbital context and its relationship to the JFC population.


\section{Dynamical Modeling of the TNO-Centaur-JFC transition}\label{sec:methods}

We use forward modeling to determine how objects evolve dynamically from TNO reservoirs, through the Centaur population, and into the JFC region. The latter transition involves a relatively transient yet robust phase of SW1-like orbits, which we call the ``JFC Gateway region.'' 

To account for vastly differing timescales in different regions of the solar system, we model this in two parts: (1) the dynamical evolution of the TNO source population into the Centaur population ($\sim10^8$--$10^9$~yr); (2) detailed evolution of Centaur orbits ($\sim10^6$--$10^7$~yr), their  passage through the JFC Gateway region, and subsequent evolution as JFCs ($\sim10^4$--$10^5$~yr). The first part can be modeled with only gravitational perturbations from the Sun and giant planets, greatly reducing the computational cost of long-timescale simulations. The second part requires including gravitational perturbations set by the terrestrial planets and thus a much smaller integration time step. 

To model Centaur production, we integrate test particles representing TNO populations forward in time for 500~Myr under the gravitational influence of the Sun and four giant planets. 
We use the Mercurius package within the {\sc rebound} orbit integration software package \citep{Rebound}, which allows us to track particles through close encounters with the planets. 
We follow each test particle until it reaches heliocentric distances either greater than 2000~au or less than 4~au. The inner boundary is set by the absence of the terrestrial planets; the outer boundary is set by the absence of galactic tides and other external influences that become important far from the Sun. As test particles enter the Centaur population (which here we define as orbits entirely contained in the giant planet region; $q>5.2$~au and $Q<30.1$~au), we record their orbital history at a sufficiently high cadence to allow us to produce high-resolution maps of their ensemble orbital evolution.

We consider two different TNO source region models for these integrations to test whether the distribution of Centaurs (and the nature of the Centaur to JFC transition) depends on the assumed source region distribution. We model: (1) the actively scattering trans-Neptunian population and (2) a simplified ``stirred'' classical Kuiper belt population.  The actively scattering objects have orbits that are currently evolving in semi-major axis due to perturbations from the giant planets \citep{Gladman:2008}. We use a scattering population model from \cite{Kaib:2011} as our initial conditions, which is consistent with current observations of the outer solar system \citep{Shankman:2016,Lawler:2018}. This model extends to the inner Oort cloud, so we take only orbits with semi-major axes smaller than 1000~au. We start with $\sim17,000$ test particles, of which $\sim1100$ enter the Centaur population within 500~Myr. 

For the classical Kuiper belt, we model the ``stirred'' population: objects confined to semi-major axes $a=$42-47~au with perihelia $q>37$~au \citep{Petit:2011}. We assume a Rayleigh distribution of inclinations with a width of $5^\circ$ (for comparison, the model scattering population's width is $\sim15^\circ$). We find that $\sim2300$ of our initial $\sim25,000$ stirred classical belt particles enter the Centaur region within 500~Myr\footnote{Note that this does not reflect the current classical belt's absolute Centaur production rate because these simplified initial conditions include unstable regions of resonances with Neptune that overlap the classical belt population.}.

\begin{figure}
\centering
\includegraphics[width=3.5in, trim=15pt 200pt 15pt 200pt, clip]{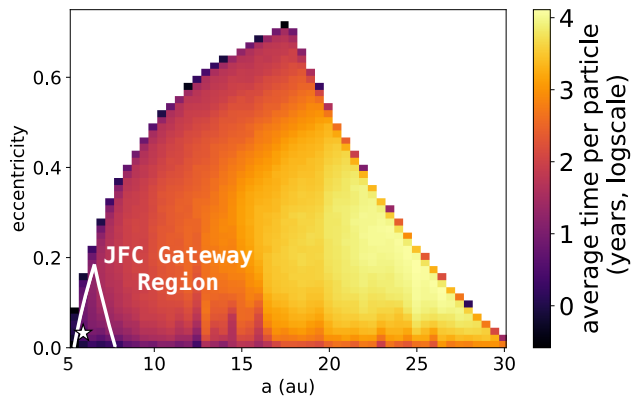} 
\caption{Average time spent (log scale; in color) per Centaur test particle at different semi-major axes and eccentricities; note that not all particles that become Centaurs evolve to small semi-major axes, so the time distribution reflects both local dynamical timescales and the probability of a particle arriving at that $a-e$ combination. SW1's location is indicated with a small star and the white lines indicate the boundaries of our JFC Gateway region.}
\label{f:time-weighted-Centaurs}
\end{figure}

As expected, highly chaotic evolution in the giant planet region largely erases these Centaurs' initial TNO conditions. Fig.~\ref{f:time-weighted-Centaurs} shows the time-weighted distribution of particles in the Centaur region (we combined both simulations because there are no significant differences in the $a-e$ distribution). The median time these test particles spend with orbits entirely within the giant planet region is $\sim2.4$~Myr, the majority of which is spent in the outer Centaur region; approximately 80\% and 50\% of Centaur test particles reach semi-major axes interior to Uranus and Saturn, respectively.

We define the ``JFC Gateway region'' as containing non-Jupiter crossing orbits with relatively low eccentricities and aphelia well-separated from Saturn: $q>5.4$~au (perihelion outside Jupiter's aphelion) and $Q<7.8$~au (aphelion $>4$ Hill radii away from Saturn's semi-major axis and $\gtrsim3$ Hill radii from Saturn's perihelion). Our simulations produced $\sim3200$ Centaurs within 500~Myr, and $\sim700$ (21\%) of these particles spend time in this JFC Gateway region. Note that this percentage and the total time these Centaurs spend in the JFC Gateway region (Fig.~\ref{f:time-weighted-Centaurs}) reflects only evolution prior to entering the inner solar system due to the absence of the terrestrial planets in these initial simulations.

To explore the transition between the Centaurs and JFCs (which requires including the terrestrial planets in the simulations), we took as initial conditions the orbit of every Centaur test particle from the above simulations at the time step that it first reached semi-major axis $a<30$~au. These orbits were each cloned 10 times by randomizing mean anomalies, longitudes of ascending node, and arguments of perihelion, and then integrated forward using Mercurius under the influence of the Sun, the terrestrial planets, and the giant planets for $10^8$~yr. Test particles were removed if they came too close to the Sun ($q<0.05$~au) or were ejected to very distant ($Q>2000$~au) orbits. Orbital histories for the test particles were output at increasingly frequent intervals as they moved to smaller semi-major axes (ranging from every 5000~yr near Neptune to every 5~yr in the inner solar system). As noted earlier, the evolution of test particles in the inner solar system does not strongly depend on the particular TNO distribution, so we combine both source regions in our analysis, giving us a sample size of nearly 12,000 JFC test particles.


\section{Results and Implications} \label{sec:results}

\begin{figure*}
    \centering
       \begin{tabular}{c c}
    \includegraphics[width=3in, trim=5pt 150pt 5pt 150pt, clip]{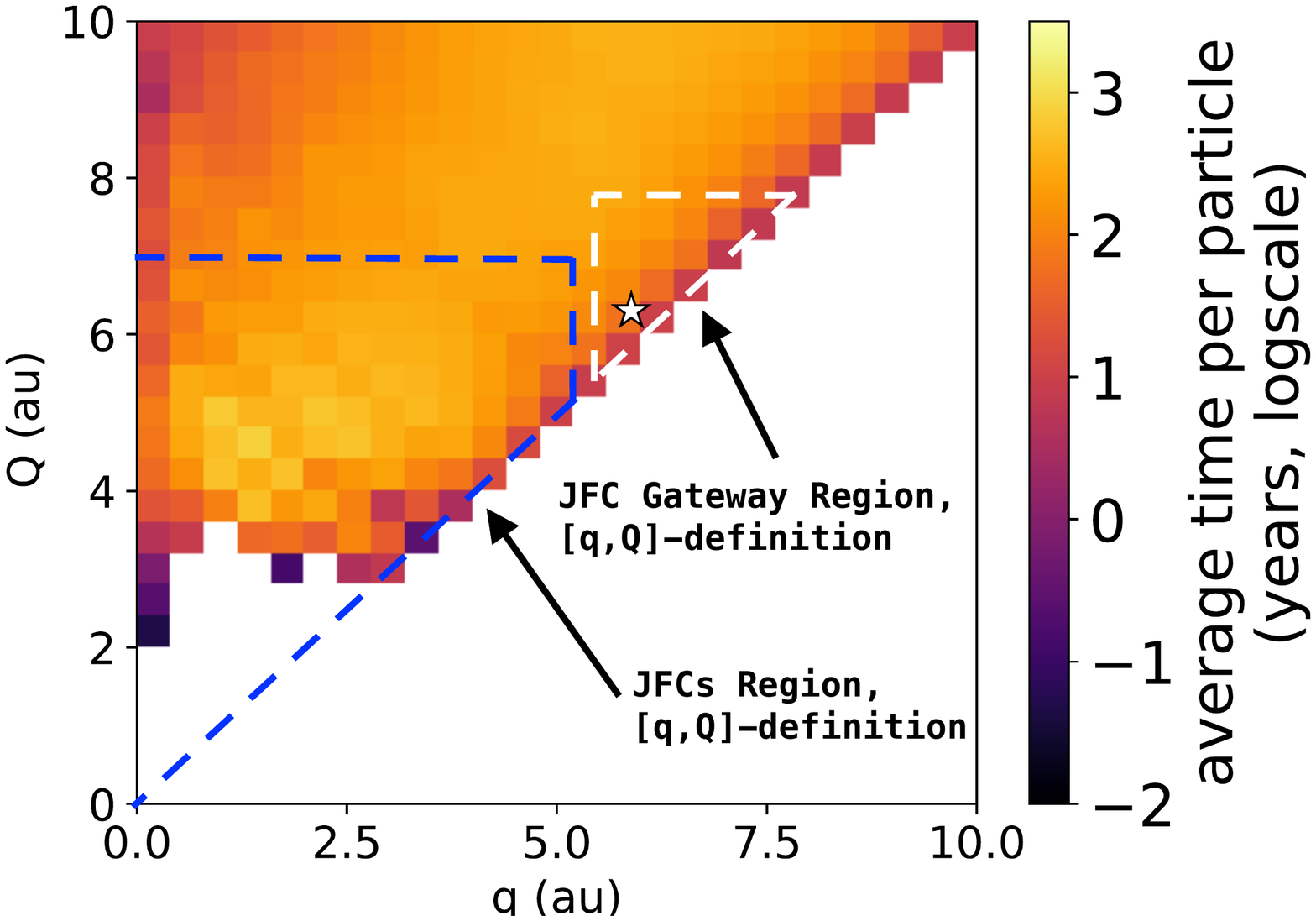} &
        \includegraphics[width=3in, trim=5pt 150pt 5pt 150pt, clip]{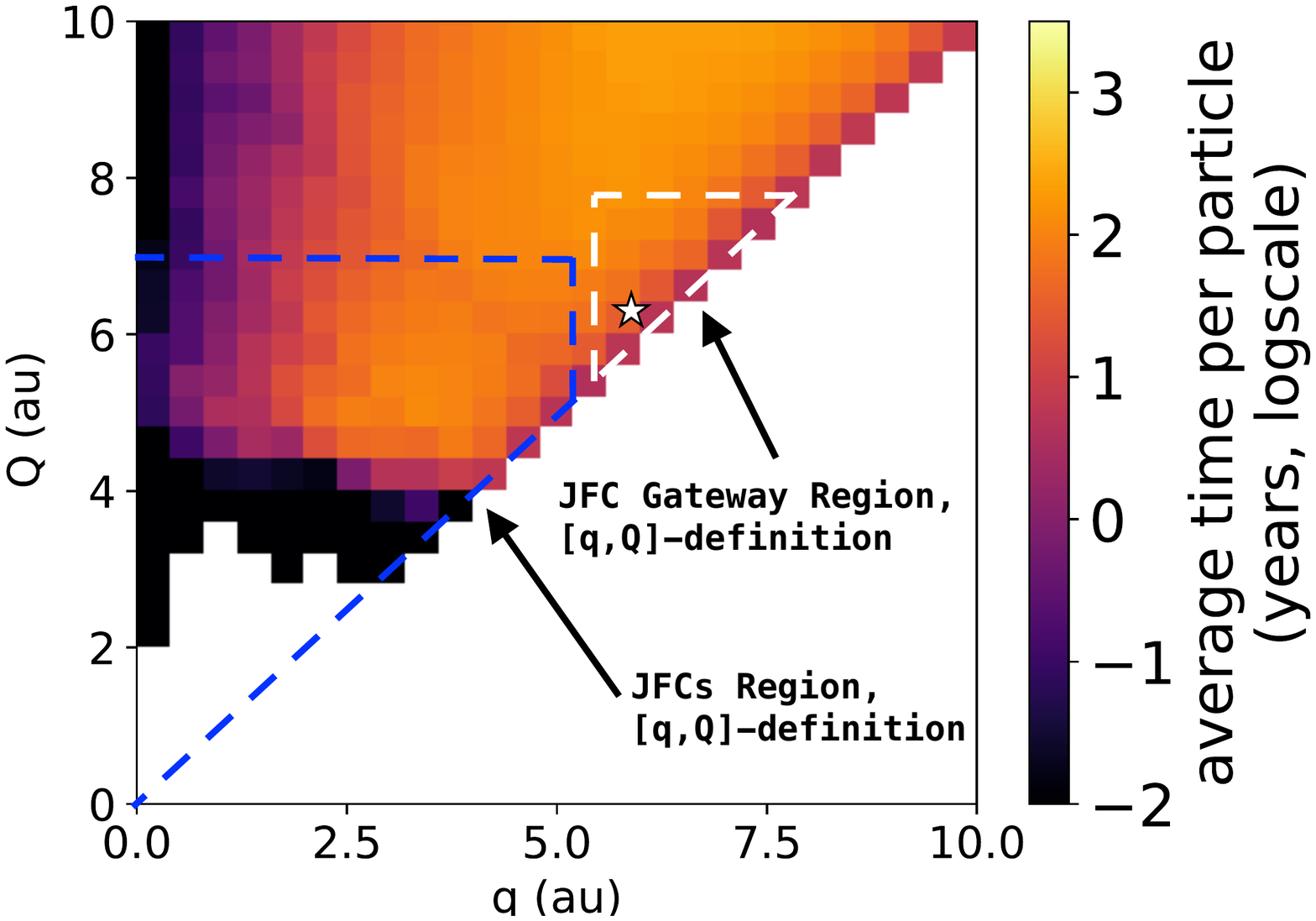}\\  
            \includegraphics[width=3in]{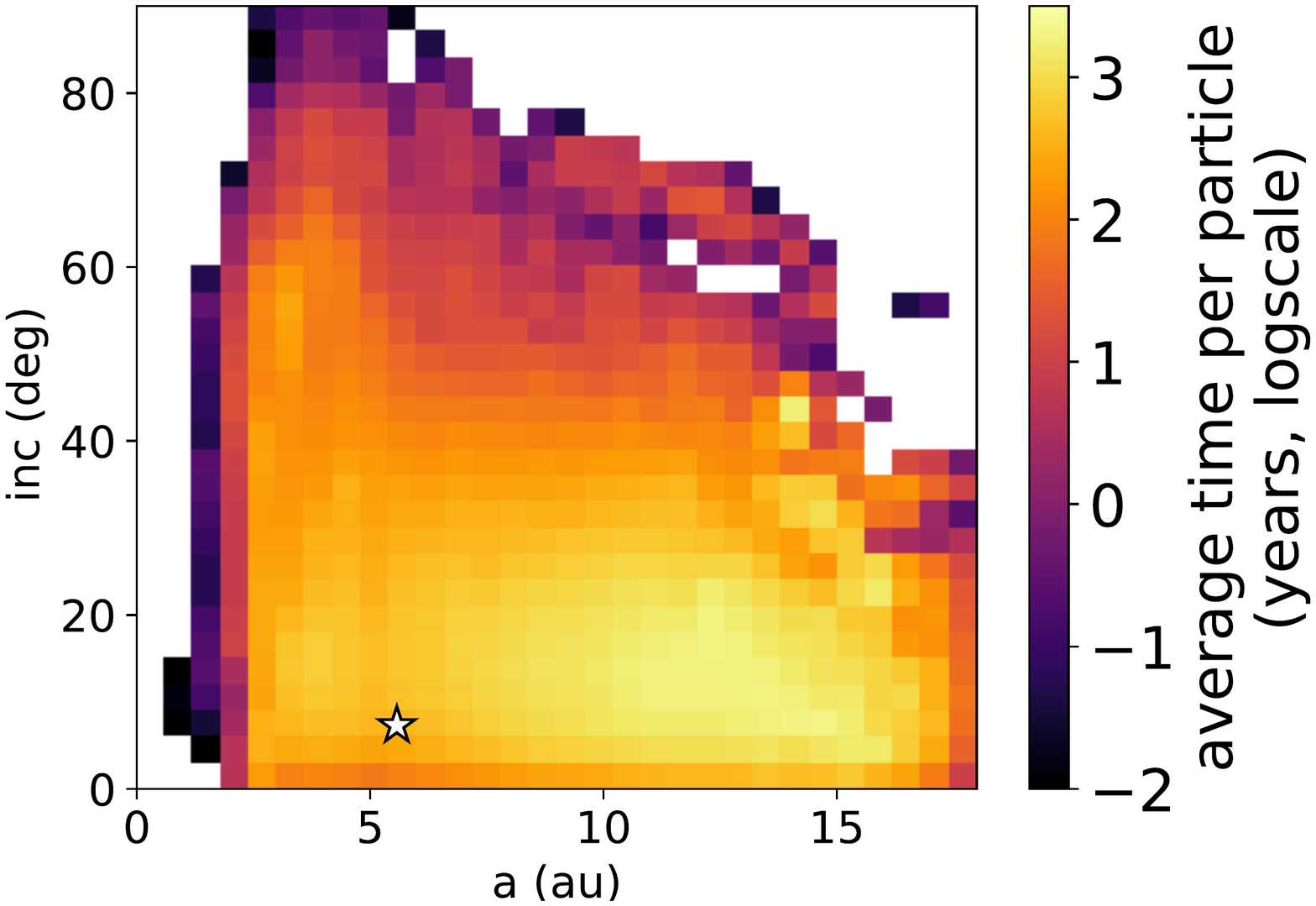} &
        \includegraphics[width=3in]{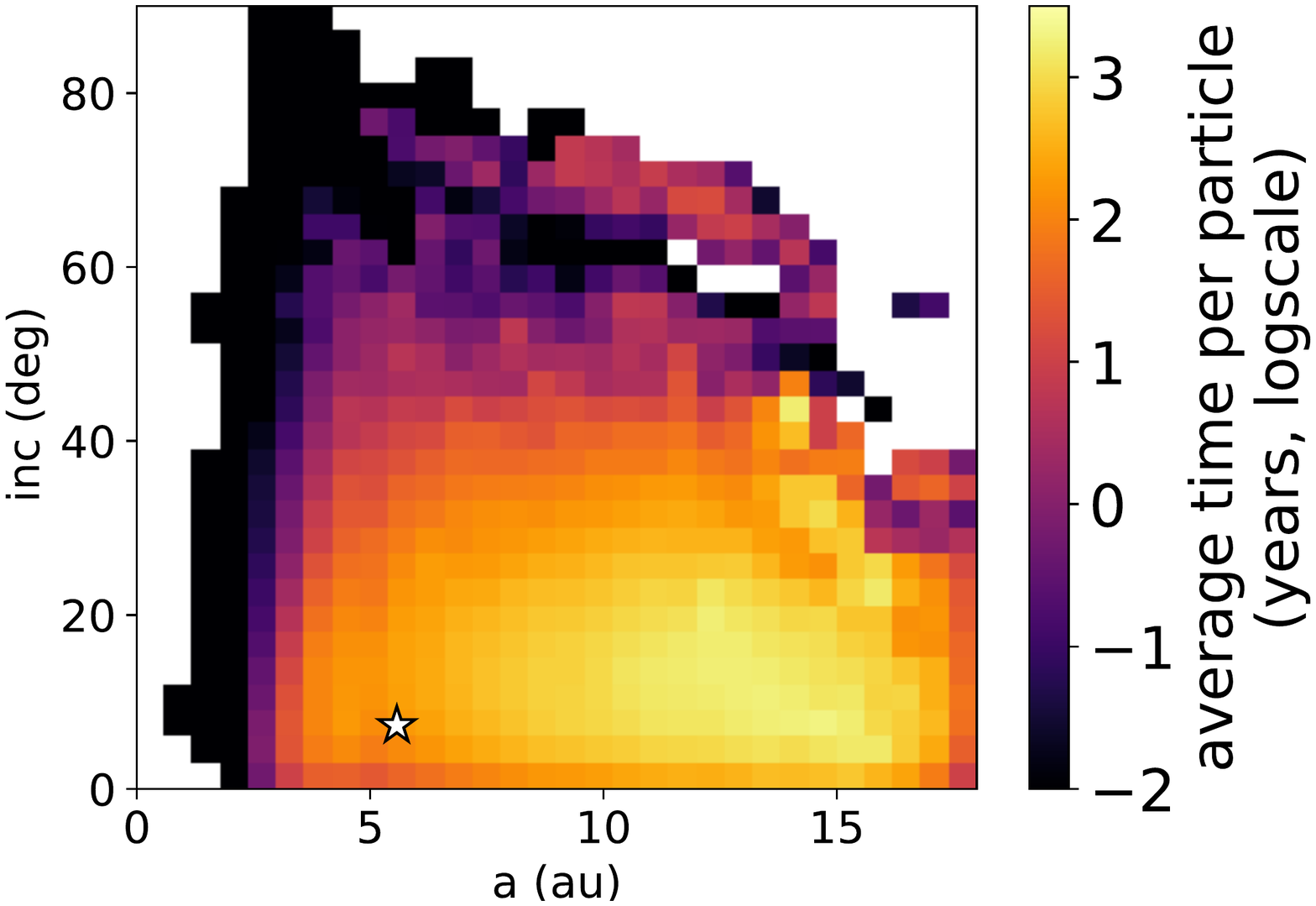}\\  
        \end{tabular}
    \caption{Time-weighted orbital distribution of JFC test particles from our full simulations (using our [q,Q]-JFC definition to select JFCs). \textbf{Left panels:} purely dynamical evolution, no fading law applied. \textbf{Right panels:} particles are weighted according to the \cite{BrasserWang2015} fading law (Section~\ref{sec:fading}). The top panels show aphelion vs. perihelion distributions near the transition between Centaurs and JFCs, with the JFC region denoted by the dashed blue line and the Gateway region denoted by the dashed white line. The bottom panels show inclination vs. semi-major axis. In each panel the white star denotes SW1's location. It is clear that fading decreases the amount of time JFCs spend at small perihelion distances and large inclinations. JFCs that fade also spend less time in the JFC Gateway region because fading limits their ability to revisit the region after spending time as inner solar system JFCs.}
    \label{fig:jfcdist}
\end{figure*}

Table~\ref{table:sims} summarizes our simulation results for the transition between the Centaurs and the JFCs, including how many JFC test particles spend time in the JFC Gateway region, their median and average residency time there, and the relative timing of Gateway occupancy. Accounting for only gravitational evolution (we consider the effects of fading in Sec.~\ref{sec:fading}), we find that 72\% of all JFCs spend some time in the Gateway region. The median time spent in this region is 1700~yr; nearly half of all JFCs that evolve to $q<3$~au (where water outgassing becomes a dominant activity driver) will spend time (median: 700~yr) in this Gateway region before reaching these small heliocentric distances. 
Tracking all test particles that pass through the Gateway, we find that 77\% of these particles are classified as JFCs at some point in the simulation. Fig.~\ref{fig:jfcdist} shows the time-weighted orbital distribution of all our JFC test particles.

\begin{table*}[]
 \caption{Summary of simulation results ([q-Q] definition for JFCs)\label{table:sims}}
 \begin{center}
\begin{tabular}{llll}
\hline\hline
\multicolumn{4}{l}{\textbf{All test particles}}   \\ \hline
\multicolumn{2}{c}{} & no fading & BW15 fading    \\
\multicolumn{2}{l}{\# JFCs generated } & 11792  & 11792  \\ 
\multicolumn{2}{l}{\# JFCs passing through JFC Gateway phase} & 8465 & 7817  \\
\multicolumn{2}{l}{Median time spent in JFC Gateway phase (yr)}  & 1750 & 425 \\
\multicolumn{2}{l}{Average time spent in JFC Gateway phase (yr)} & 8000 &  1675 \\ \hline
\hline
\multicolumn{4}{l}{\textbf{Particles having a JFC Gateway phase prior to reaching}}\\
\hline
Perihelion distance & Percentage &  $t_{median}$ (yr)& $t_{average}$ (yr) \\ 
$q<4$~au &  38\% &  625 & 4050 \\ 
$q<3.5$~au & 44\% &   650 & 3950 \\ 
$q<3$~au & 49\% &  700 & 3925 \\ \hline
 \end{tabular}
 \end{center}
 \tablecomments{The `no fading' columns in the top portion of the table reflect purely gravitational  evolution in the simulations. The `BW15 fading' column reflects those same simulations with \cite{BrasserWang2015} empirical fading law applied as a weighting factor for each particle. The bottom portion of the table details the relative timing of the JFC gateway phase. For example, 38\% of JFCs visit the Gateway before reaching $q<4$~au, with a median residence time of 625~yr prior to reaching $q<4$~au.}
\end{table*}

\subsection{Evolving back and forth between the Centaurs and JFCs}\label{sec:fading}

Gravitational perturbations can evolve JFC nuclei back into the Centaur population, complicating our characterization of the Gateway region and comparisons between our dynamical study and observations of JFCs and Centaurs. For example, simulations of only the initial Centaur evolution (Sec.~\ref{sec:methods}, Fig.~\ref{f:time-weighted-Centaurs}), indicate 21\% of Centaurs enter the JFC Gateway region. However, in our full simulations of inner and outer solar system orbital evolution (Fig.~\ref{fig:jfcdist}), this increases to $\sim30$\%.

Some of the test particles in the JFC Gateway region arrive there after being at small heliocentric distances (see Table~\ref{table:sims}). In the real solar system, this would require comet nuclei to survive the thermal environment of its residency in the inner solar system where surface and near-surface volatile material (e.g. water ice) is heated and can quickly sublimate and trigger mass loss \citep[e.g.][]{Meech:2004CometsII, Guilbert-Lepoutre2015}. In addition to sublimation, other mechanisms can affect the physical evolution of comet nuclei, such as lag deposit insulation that quenches activity \citep[e.g.][]{Prialnik:2004CometsII}, sublimative torques that can spin up nuclei and trigger mass wasting or splitting \citep[e.g.][]{SteckloffJacobson2016,SteckloffSamarasinha2018}, and localized erosion that can induce mass loss \citep[e.g.][]{Vincent2017}. Thus,  over time nuclei can be removed from the comet population. 

We must account for such physical processing in our simulated population statistics.
The exact physical evolution for any particular comet is complicated and stochastic. However, the physical processing of an ensemble of objects can be approximated using a ``fading law'', which describes the typical behavior of comet nuclei based on fits to observed trends. 

There are different ways to construct a JFC fading law, depending on what mechanisms are considered most prominent \citep[e.g.,][]{ChenJewitt1994, Boehnhardt:2004CometsII, Samarasinha:2007, Belton2015}. For our purposes, we chose the \citet{BrasserWang2015} implementation, which is derived from a large set of dynamical simulations fitted to the observed JFC population. Their best-fit result is a delayed power law that describes how the visibility $\phi_m$ of a comet is reduced over time as a function of the number of perihelion passages $m$ it experiences with $q<2.5$~au: 
\begin{equation}
    \phi_m \propto (M^2 + m^2)^{-k/2},
\end{equation}
where $k\sim1.4$ and $M=40$. \cite{BrasserWang2015} analysis included comets with diameters down to 2.3~km. This size limit is sufficient for our purposes, as it is about half the diameter of the smallest estimate for an observed object in the Gateway region (Sec.~\ref{sec:intro}). Comet nuclei size distributions for smaller sizes are currently poorly constrained and incomplete \citep{Snodgrass2011, Fernandez2013, Bauer2017}, and the fragmentation rate at such small sizes is likely very high \citep[see][]{Belton2015}.     

To apply this fading law to our simulations, we use $\phi_m$ to weight each test particle in the analysis of the time-averaged population. With this fading law, we still find that the majority of JFCs (66\%) spend time in the Gateway region. The median total time in the Gateway region drops from 1750~yr to 425~yr (Table~\ref{table:sims}), because JFCs reaching low perihelion distances tend to fade before moving back out into this region. Fig.~\ref{fig:jfcdist} shows the time-weighted distributions of our JFC test particles with and without this fading law.  In particular, the inclination vs. semi-major axis distribution highlights why fading laws are typically implemented. Purely dynamical interactions with the planets tend to increase orbital inclinations as test particles spend more time in the inner solar system, making their distribution inconsistent with the lower observed inclinations in the real solar system \citep[see, e.g.,][]{Nesvorny2017}; fading limits particle lifetimes and thus limits their inclinations.


\subsection{Expected occupancy of the JFC Gateway region} \label{sec:populations}

The presence of this Gateway region, through which the majority of JFCs pass, provides a testable connection between the observed JFC and Centaur populations.
We can estimate the time-averaged number of objects currently residing in this region beyond Jupiter at different sizes as a check on the validity of our dynamical model. To do this, we employ a Centaur size distribution function based on the observed cratering record in the Pluto-Charon system \citep{Singer2019}, the observed size distribution function of JFCs with measured radii $r>4$~km \citep{Snodgrass2011}, and the fragment size fit for C/1999 S4 (LINEAR) following its total breakup \citep{Makinen2001}. From these studies, we obtain a power-law size distribution for the Centaurs:
\begin{equation}
dN_{r}=-k{\alpha}r^{-(\alpha+1)} dr, 
\end{equation}
where $\alpha=3$ and $k=6.5\times10^6$.
This predicts $\sim50$  Centaurs with $r>50$~km, which is in reasonable agreement with the estimated intrinsic Centaur population at these sizes. \cite{Lawler:2018} estimates there are $110^{+60}_{-40}$ Centaurs with $H_r<8.66$. Assuming an albedo range of 4-12\% \citep[e.g.,][]{Bauer2013}, this translates to $\sim40^{+20}_{-15}$ to $110^{+60}_{-40}$ Centaurs with $r>50$~km.
Integrating over all radii $r>1$~km, we obtain a total Centaur population of $6.5\times10^6$, which is consistent with previous estimates \citep{Sheppard2000}.  

The expected time-averaged occupancy of the Gateway region for specific size ranges is obtained by multiplying the integrated size distribution by our derived values of: the percentage of all Centaurs that enter the Gateway region (30\%, including JFCs re-entering the Centaur population);  the ratio of the median residency time in the Gateway (Table~\ref{table:sims}, depends on fading) to the  median dynamical lifetime in the Centaur region ($2.6\times10^{6}$~yr, Sec.~\ref{sec:methods}). If we consider the size range of the four objects observed to be in the Gateway ($2<r<32$~km), the expected occupancy rate from our model ranges from $\sim5$ (with fading) to $\sim20$ (without fading) objects, consistent with observations. Objects as large or larger than SW1 (radius estimated to be $23<r<32$~km; \citealt{Bauer2013,Schambeau2018}) are significantly less likely, with an expected occupancy rate ranging from 0.02 (with fading) to 0.085 (without fading). That makes large objects in the Gateway region rare enough on the $\sim10{^4}$~year timescale of human civilization that SW1 is likely the first such object to enter the Gateway in recorded history.   

An interesting prediction from our model is the large expected occupancy rate in the Gateway region for Centaurs with $r>1$~km, ranging from $\sim300$ (with fading) to $\sim1000$ (without fading). Such a large instantaneous population would experience very close encounters with Jupiter. This may explain the impact frequency of events such as Shoemaker--Levy 9 (1993) and the later, serendipitous impact detection in 2009 \citep{SanchezLavega2010, Hueso2013}.  


\subsection{Implications for SW1} \label{sec:sw1}

SW1 falls into the category of objects that are large enough to be relatively unaffected by physical fading mechanisms \citep[e.g.][]{Belton2015}. These larger objects are thus more likely than small objects to revisit the Gateway region after spending time at low perihelion distances. Here we discuss SW1's activity characteristics and likely orbital history in this context.

\begin{figure}
    \centering
    \includegraphics[width=3.5in]{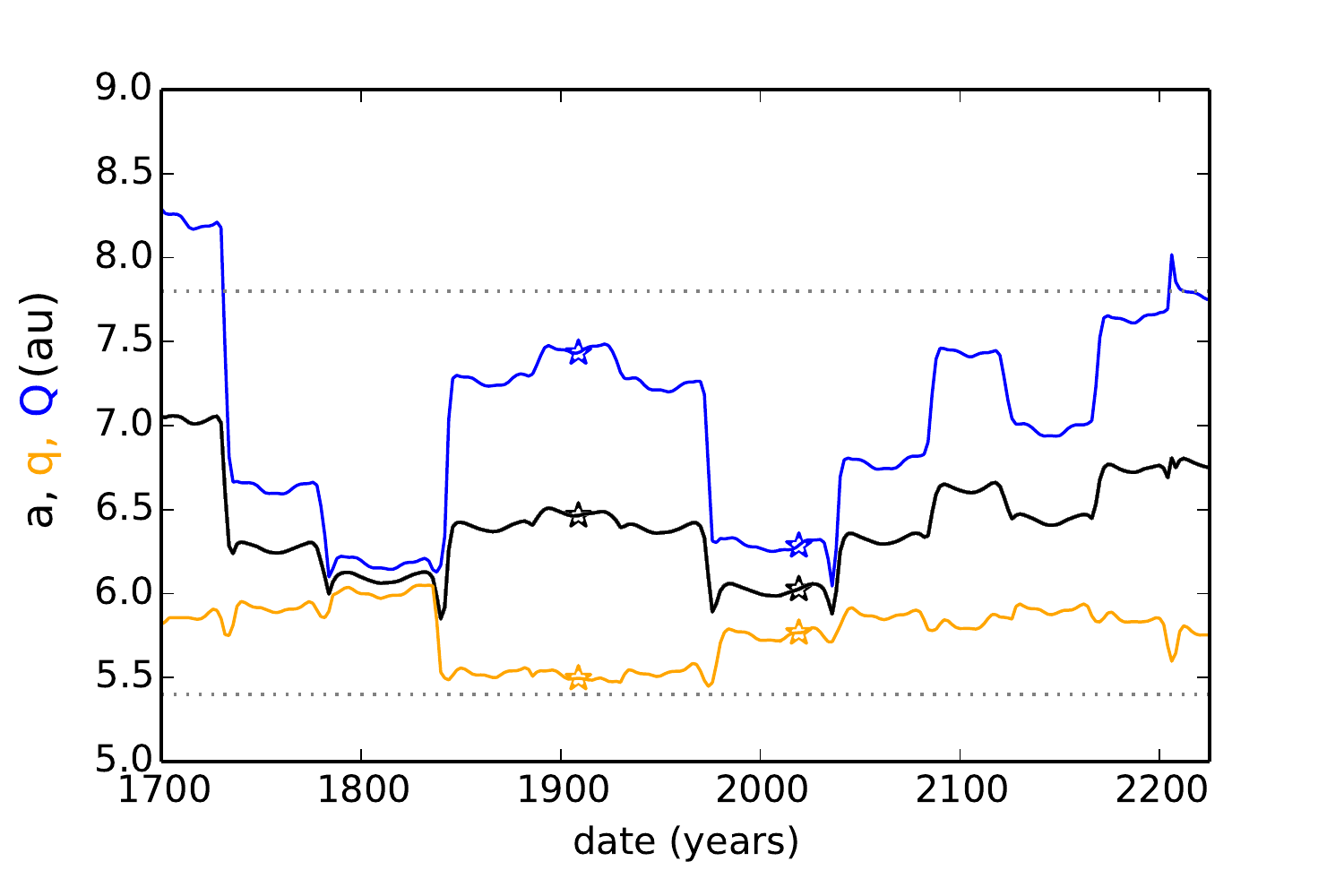}
    \caption{Short-term orbital evolution of SW1's semi-major axis $a$ (black line), perihelion $q$ (gold line), and aphelion $Q$ (blue line). The dashed horizontal lines denote our JFC Gateway region. SW1's orbit significantly changed due to a Jupiter conjunction in 1975, and it will change again soon in 2038 due to another conjunction with Jupiter. The stars show SW1's observed orbits in 2019 and in 1908.}
    \label{fig:sw1-orbit}
\end{figure}

Due to its close proximity to Jupiter, SW1's orbit evolves significantly on short timescales. Fig.~\ref{fig:sw1-orbit} shows the recent past and near future evolution of SW1.\footnote{We take the Jan 2010 orbit fit and uncertainties from JPL Horizons \citep{jplhorizons} as our initial conditions.} 
Currently, SW1's orbit undergoes semi-major axis and eccentricity changes when it comes to conjunction with Jupiter every $\sim50$~yr. Its present-day, very-low-eccentricity orbit ($e=0.043$) was established after a 1975 conjunction and will continue until a 2038 Jupiter conjunction nearly doubles its eccentricity and pushes its semi-major axis out to its current aphelion, causing SW1 to experience much wider variations in solar heating. It will be interesting to see whether this new orbit affects its level of activity and cycle of outbursts.

As Fig.~\ref{fig:sw1-orbit} shows, the backward integrations reproduce SW1's observed orbit more than 100~yr ago,\footnote{The minor planet center orbit fit for the epoch 1908 October 26.}
providing a lower bound on how long from the present epoch the integrations (which neglect non-gravitational forces) reliably predict SW1's orbit. Integration of many clones of SW1 (within its orbital uncertainties) indicate that SW1 can be tracked back approximately $\sim700$~yr before the clones diverge as a result of chaos and a particularly strong Jupiter encounter.

From these calculations, we can only conclusively say that SW1 has spent at least a few hundred years in the JFC Gateway region. We note that SW1's current low inclination ($\sim10^\circ$) is suggestive of it not having spent significant time in the inner solar system (see bottom panels of Fig.~\ref{fig:jfcdist}). SW1's clones can be integrated forward for several hundred years before their evolution diverges. Further into the future we have only a statistical description of SW1's evolution: in the next $10^4$~yr, it has a $\sim65\%$ chance of becoming a JFC (using our [q-Q]-definition).

We note that based on SW1's size and activity pattern \citep[somewhat similar to Hale--Bopp, see][]{Wierzchos2017, WomackDistantReview2017}, were it to enter the JFC population it would be the largest short-period comet ever recorded and most likely the brightest short- or long-period comet ever seen. We have no historical record of such an apparition and no object close to SW1's size is recorded in the observed short- or long-period comet populations. This implies that Centaur-to-JFC transitions in this size range are rare on the timescale of human civilization and occur with a frequency smaller than the dynamical lifetime of the typical JFC. Interestingly, our dynamical model (Table~\ref{table:sims}; Section~\ref{sec:populations}) supports this conclusion, without setting SW1's observability or active state as imposed constraints. 

While SW1's exact long-term orbital history is unknown, we can say that it is more consistent with it being a new visitor to the Gateway region, rather than a return visitor that had ever spent any significant time in the inner solar system. This conclusion is supported by SW1's current low-inclination orbit ($<$$10^\circ$), high activity levels comparable to dynamically new long-period comets at the same heliocentric distance, and the lack of a historical record of a short-period ``super-comet''.


\section{Conclusions} \label{sec:conclusions}

Our dynamical study of the transition between Centaurs and JFCs, motivated by the intriguing properties of SW1, enabled us to identify and characterize an important orbital niche -- a Gateway to/from the JFC population. This region of low-eccentricity orbits just exterior to Jupiter ($q>5.4$~au, $Q<7.8$~au) is currently populated by four known objects, of which SW1 is the most prominent in terms of size, activity, and observational record. 

Our dynamical models reveal that the vast majority of objects  (77\%) in this Gateway region will become or have already been JFCs, and the majority of JFCs (66-72\%, depending on fading) pass through this Gateway. Approximately half of all JFCs will pass through this Gateway region before experiencing significant water-driven sublimation ($q$$<$3~au).

The dynamical importance of the JFC Gateway region in which SW1 resides places its activity behavior in much clearer context as a transition between other kinds of active objects in the outer solar system. This can serve as a complimentary dynamical framework for studies of outer solar system activity and surface color distributions \citep[e.g.][]{Jewitt2009, Jewitt:2015}. 
In this context, SW1's active state, likely driven by a non-water--ice sublimation process, is probably typical of the early significant evolutionary processing of objects that subsequently become JFCs. 

Our identified Gateway region is an important, albeit brief, stage of orbital, and thus likely physical, evolution that appears exceptionally common for Centaurs to pass through. 
It deserves additional studies in both modeling and observational capacities to better understand the physical processing of nuclei surfaces and sub-surfaces as objects transition in and out of the Gateway region and better constrain the number and size of such objects. 
While the pathway SW1 is taking is very common, its short-lived nature means it is rare for an object as large as SW1 to be present at any given epoch. This adds to the motivation for studies of SW1 in particular.


\acknowledgements{An allocation of computer time from the UA Research Computing High Performance Computing (HPC) at the University of Arizona is gratefully acknowledged. G.S. acknowledges support from NSF grant 1615917 and NASA award 80NSSC18K0497. K.V. acknowledges support from NASA grants NNX15AH59G and 80NSSC19K0785 and NSF grant AST-1824869. J.S. acknowledges support from NSF grant 1910275 and NASA award 80NSSC18K0497. M.W. acknowledge support from NSF grant 1615917. The authors acknowledge useful conversations with Y.R. Fernandez and suggestions from the referee.}  


\bibliography{sample}


\end{document}